\documentclass[prb,superscriptaddress,floatfix,twocolumn,showpacs,amsfonts,amsmath,amssymb]{revtex4}
\usepackage{graphicx} 
\usepackage{dcolumn} 
\usepackage{bm} 
\usepackage{color}

\usepackage{tikz}

\begin{document}

\title{Interference in spin-orbit coupled transverse magnetic focusing; emergent phase due to in-plane magnetic fields}

\author{Samuel Bladwell}
\affiliation{School of Physics, University of New South Wales, Sydney 2052, Australia}
\author{Oleg P. Sushkov}
\affiliation{School of Physics, University of New South Wales, Sydney 2052, Australia}

\begin{abstract}

Spin-orbit (SO) interactions in two dimensional systems split the Fermi surface, and allow for 
the spatial separation of spin-states via transverse magnetic focusing (TMF). In this work, 
we consider the case of combined Rashba and Zeeman interactions, which leads to a Fermi 
surface without cylindrical symmetry. While the classical trajectories are effectively unchanged, 
we predict an additional contribution to the phase, linear in the applied in-plane magnetic field. We show that
this term is unique to TMF, and vanishes for magnetic (Shubnikov de Haas) oscillations. Finally 
we propose some experimental signatures of this phase.

\end{abstract}

\maketitle

\section{Introduction}

Transverse magnetic focusing (TMF) has a long history, 
being employed in metals and semi-conductors, and has
been used to investigate the shape of the Fermi surface\cite{Sharvin1965, Sharvin1965a, Tsoi1974, Tsoi1999, Vanhouten}.
 A TMF experiment consists of a source and a detector, separated by a distance $l$, 
with charges focused from the source to the detector via a weak transverse magnetic field. 
It is the direct translation of charge mass spectroscopy to the solid state. 
Despite the nearly half century of 
experimental history, TMF is still producing novel results, 
with the most recent application in systems with non-quadratic 
dispersion relations, such as Graphene\cite{Taychatanapat2013} 
and two dimensional charge gases with large spin-orbit (SO) interactions\cite{Rokhinson2004}.
In spin-orbit coupled systems, the spin-split Fermi surfaces result 
in a ``doubled" focusing peak, which provides a 
novel platform investigations of polarisation effects in the source 
and detector quantum point contacts\cite{Usaj2004, Rokhinson2006}. 
The separation of the peaks also allows for the direct determination 
of the magnitude of the spin-orbit splitting, hence 
TMF can be used in addition to quantum magnetic oscillations to yield detailed 
information about spin-orbit coupled electron and 
hole systems.

Much of the theoretical and experimental work concerning 
TMF with large spin-orbit splittings has considered 
a singular dominant SO interaction. This leads to a cylindrically 
symmetric Fermi surface, and a double peak structure
that is, in essence, two copies of the single peak structure\cite{Zulicke2007}. 
This assumption is well justified for 
many typical experimental systems grown along high symmetry
 crystal axes, as classical trajectories are not 
significantly altered except in the case of extremely 
large asymmetry\cite{Bladwell2015}. While a sufficiently large 
secondary SO interaction can lead to magnetic breakdown
 like behaviour\cite{Reynoso2007}, the requirement
for resolution of the double peak structure means that the
 typical regime is one characterised by the secondary SO 
interaction being weaker than the primary 
interaction that yields the double peak structure of spin-split
TMF. 

Like earlier studies in semiconductors, SO coupled systems 
have Fermi wavelengths comparable to the 
feature size making interference an important feature of the 
magnetic focusing spectrum\cite{Vanhouten, Bladwell2017}. With the
addition of SO coupling, the interference effects are 
further enriched, and yield new methods of studying
SO interactions. In this paper, 
we focus on the problem of interference in TMF in systems 
with non-cylindrically symmetric spin-orbit interactions due to an 
applied in plane field. 
Importantly, due to the
large pre-factors, proportional to the Fermi momentum and
 focusing length, relatively small in plane fields can 
 lead to large phase contributions.
While the classical trajectories are effectively unchanged, an additional
phase term emerges, linear in the applied magnetic field. We show that this additional 
contribution to the phase can significantly alter the TMF interference spectrum.

Our paper is organised as follows. In Sec. \ref{trajs} we present the 
classical trajectories for magnetic focusing, and introduce the 
relevant Hamiltonian. Following on from this, in Sec. \ref{interference} 
we develop a theory of interference in the absence of 
cylindrical symmetry, building on previous work on interference in 
TMF with SO coupling\cite{Bladwell2017}. 
 Finally, in Sec. \ref{discuss} 
 we consider some relevant examples, with a minimalistic 
 model of the injector and detector wave functions.


\section{Spin-orbit interactions and classical trajectories}
\label{trajs}

Semiconductor heterostructures allow for a great diversity of SO interactions. While the approach 
we will detail is general, for specificity we will consider two interactions; the Rashba interaction
resulting from a lack of surface inversion symmetry in the sample, and the Zeeman interaction due to 
an applied in plane magnetic field. These two interactions have the advantage of being tunable. 
In electron systems, the Rashba interaction has the kinematic structure\cite{Bychkov1984},  
\begin{eqnarray}
{\cal H}_{R, e} = i \frac{\gamma_1}{2} p_- \sigma_+ +h. c. 
\label{spinorbit1b}\\
\sigma_\pm  = \sigma_x \pm i\sigma_y \quad p_\pm = p_x + ip_y \nonumber
\end{eqnarray}
where $\gamma_1$ is material parameter dependent on the electric field perpendicular to the 
two dimensional plane. The Pauli matrices $\sigma$ correspond to electron spin $s = 1/2$, and the selection
rule for $\sigma_\pm$ is $\Delta s_z  = \pm 1$. 
Spin splitting in the magnetic focusing spectrum was recently observed in InGaAs quantum wells\cite{Lo2017}.
In GaAs heterostructures, the spin-orbit interaction is typically not large enough to obtain a spin-split magnetic focusing 
spectrum. Heavy hole gases can also be engineered to have a Rashba spin orbit interaction\cite{Winkler2003}. Due 
to the heavy holes having angular momentum $J_z = \pm 3/2$, the Rashba interaction arises from the combined action
of both the Luttinger, $({\bf p \cdot J})^2$, and Rashba terms, ${\bf p} \cdot ({\bf J\times z})$, with
${\cal H}_R   \propto ({\bm J}\cdot {\bm p})^2({\bm p}\cdot{\bf J \times z })$. Typically, the light holes
$J_z = \pm 1/2$,
have significantly higher energy, and it is more convenient to work in the subspace spanned 
by the Pauli matrices, with $J_\pm^3 \rightarrow \sigma_\pm$. The selection rule is $\Delta J_z = \pm 3$. 
 In this subspace the kinematic 
structure is\cite{Winkler2003}
\begin{eqnarray}
{\cal H}_{R, h} = i \frac{\gamma_3}{2} p_-^3 \sigma_+ +h. c.
\label{spinorbit1}
\end{eqnarray}
where $\gamma_3$ is a material parameter analogous to $\gamma_1$.

 To induce an asymmetry in the spin-split Fermi surface, we consider an applied in plane magnetic field. For electrons, this results in
  the usual 
 Zeeman interaction, 
 \begin{eqnarray}
{\cal H}_{Z, e} = \frac{g}{2}\mu_B  B_- \sigma_+ +h. c.
\label{spinorbit2b}
\end{eqnarray} 
where $g$ is the electron $g$ factor, and $B_\pm = B_x \pm iB_y$. 
 There is 
no equivalent expression for heavy holes, as $J_z = \pm3/2$ cannot be coupled directly, but 
requires the combined action of Zeeman, ${\bf J \cdot B}$ and Luttinger, $({\bf p \cdot J})^2$,
 with ${\cal H}_{Z, h} \propto ({\bm J}\cdot {\bm p})^2({\bm J}\cdot{\bm B})$.
The kinematic structure is
\begin{eqnarray}
{\cal H}_{Z, h} = \frac{g_1}{2} \mu_B p_-^2 B_- \sigma_+ +h. c.
\label{spinorbit2}
\end{eqnarray}
where we use the aforementioned subspace of heavy holes\cite{Li2016, Miserev2017}.

We use a dimensionless form of the coefficients $\gamma_3$ in Eq.(\ref{spinorbit1})  and 
$\gamma_1$ in Eq.(\ref{spinorbit1b}),
\begin{eqnarray}
&&\gamma_1={\tilde \gamma_1}\frac{\epsilon_F}{k_F} \\
&&\gamma_3={\tilde \gamma_3}\frac{\epsilon_F}{k_F^3}\nonumber \\ 
&&k_F=\sqrt{2 m \epsilon_F} \nonumber
\label{gm}
\end{eqnarray}
where $\epsilon_F$ is the Fermi energy (chemical potential).
The dimensionless coefficient ${\tilde \gamma_{1, 3}}$ represents
the value of the SO interaction at $p=k_F$ in units of the Fermi energy.
This can be directly related to the splitting the ``double" TMF peaks. 
For the heavy holes Rashba interaction, $\tilde \gamma_3$ can  as large as 
$|{\tilde \gamma_3}| \sim 0.1-0.2$, in GaAs depending on the $z$ confinement\cite{Winkler2003}.
 For the electron Rashba interaction, in InGaAs quantum wells, $\tilde \gamma_1 \sim 0.2$\cite{Lo2017}. 
 For the Zeeman interaction in holes, we consider the dimensionless coefficient, $\tilde g_1$,
 \begin{eqnarray}
&&{\tilde g}_1={g_1}{k_F^2} 
\end{eqnarray}
For GaAs heavy hole quantum wells, $\tilde g_1 \sim 1$\cite{Li2016, Miserev2017}. The electron $g$ factor in InGaAs quantum 
wells is $g\sim -9$\cite{Simmonds2008}.

We can consider the SO interaction as a momentum dependent effective Zeeman
magnetic field, ${\cal B}({\bf p})$. Hence the Hamiltonian is
\begin{eqnarray}
{\cal H} = \frac{\bf p^2}{2m} + {\cal B}({\bf p}) \cdot \boldsymbol{\sigma} \\ \nonumber
\boldsymbol{\sigma} \cdot {\cal B}({\bf p}) = {\cal{H}}_{R} + {\cal H}_{Z}
\label{spinorbit3}
\end{eqnarray}
with the application of a transverse magnetic field, $B_z$, ${\bf p \rightarrow \boldsymbol{\pi} = p -} e{\bf A}$, 
with the vector potential chosen in an appropriate gauge, ${\bf A} = B_z(0, -x, 0)$. 
The semi-classical dynamics of the charge carriers are characterised by cyclotron orbits\cite{Bladwell2015}, 
with a cyclotron radius, $r_c = k_F/eB_z$, and a cyclotron frequency, $\omega_c = eB_z/m$.
Due to the curvature of the trajectories, the effective magnetic field, ${\cal B}$ evolves in time. 
Since TMF experiments are typically performed at relatively small transverse 
magnetic fields, $B_z \le 0.1$T, the spin adiabatically follows the effective magnetic 
field, ${\cal B}$. Provided $|{\cal B}| \gg \omega_c$, there is no tunnelling 
between the two spin states. We note that this is also a condition for a
``double" focusing peak. 

We are now in a position to explore the semiclassical dynamics. The Hamiltonian, with a 
applied magnetic field, ${\bf B}$ is
\begin{eqnarray}
&{\cal H} = \frac{\boldsymbol{\pi}^2 }{2m} + \boldsymbol{\sigma} \cdot{\cal B} \\
&{\bf B} = (B_{||} \cos\varphi, B_{||} \sin\varphi, B_z)  \nonumber \\
& \boldsymbol{\pi} = {\bf p} - e {\bf A} \nonumber
\end{eqnarray}
where $\varphi$ is the field angle, $B_{||}$ is the in plane magnetic field, and
$B_z$ is the (weak) transverse focusing field. We stress that typically $B_z \sim 0.1$T, 
while $B_{||}$ can be a few Teslas for heavy hole quantum well in GaAs\cite{Rokhinson2004}.
For electrons in InGaAs, $B_{||} \sim 1$T due to the much large $g$ factor in these systems.  
If the spin follows the effective field adiabatically, $\boldsymbol{\sigma} \rightarrow s {\cal B} /|{\cal B}|$, 
where $s$ is a pseudo scalar, and describes the two possible spin states.
The resulting adiabatic
Hamiltonian is
\begin{eqnarray}
{\cal H}_{cl} =  \frac{\boldsymbol{\pi}^2 }{2m} + s |{\cal B}|
\label{classham}
\end{eqnarray}
The semiclassical dynamics of this hamiltonian has been found with 
expansion in powers of $|{\cal B}|/\varepsilon_F$\cite{Bladwell2015}. This 
is valid in the regime $\omega_c \ll |{\cal B}| \ll \epsilon_F$. 
The effective magnetic field, ${\cal B}$, is
\begin{eqnarray}
\label{bb}
&&|{\cal B}|=\epsilon_F|{\tilde \gamma_3}| b(\theta)\\
&&b(\theta)= \rightarrow\nonumber\\
&&=\sqrt{1+2({\tilde g_1 \mu_B}/{\tilde \gamma_3 \varepsilon_F})B_{||}\cos(\theta-\varphi) 
+({\tilde g_1 \mu_B }/{\tilde \gamma_3\varepsilon_F})^2B_{||}^2} \nonumber
\end{eqnarray}
For holes, and 
\begin{eqnarray}
\label{bb}
&&|{\cal B}|=\epsilon_F|{\tilde \gamma_1}| b(\theta)\\
&&b(\theta) = \rightarrow \nonumber \\
&&=\sqrt{1+2({ g \mu_B}/{\tilde \gamma_1\varepsilon_F})B_{||}\cos(\theta-\varphi)
+({ g \mu_B}/{\tilde \gamma_1 \varepsilon_F})^2 B_{||}^2} \nonumber
\end{eqnarray}
for electrons. Evidently, these effective magnetic fields are identical, and the 
dynamics of electrons and holes are the same in this adiabatic semiclassical approach, despite 
the kinematic structure of the spin-orbit interactions, Eqs. \eqref{spinorbit1} and \eqref{spinorbit1b}
being markedly different. For clarity, in the following calculations, we will exclusively refer to  
holes. 

\begin{figure}[t!]
 \label{int1} 
  \begin{center}
    {\begin{tikzpicture}
    \draw [black, thick] (-2.5, 1.5) -- (-1.5, 1.5);
    \draw [black, thick] (-1, 1.5) -- (1, 1.5);    
    \draw [black, thick] (1.5, 1.5) -- (2.5, 1.5)node [black] at (1.2, 1.3) {source};    
      \draw [black, <->, thick] (-1.3, 1.0) -- (1.3, 1.0) node [black] at (0, 1.2) {$l$}; 
      \draw [black, ->, thick] (1.2, 1.5) -- (1.2, 3.0) node [black] at (1.2, 3.1) {$x$} ;
       \draw [red,thick,->, domain=0:180] plot ({1.2*(1 + 0*sin(\x))*cos(\x)}, {1.5 +  1.2*(1 + 0*sin(\x))*sin(\x)});
        \end{tikzpicture}}
            {\begin{tikzpicture}
 
      \draw [black, ->, thick] (1.2, 1.3) -- (1.2, 3.0) node [black] at (-0.05, 1.7) {$y$};
      \draw [black, ->, thick] (1.4, 1.5) -- (0.0, 1.5) node [black] at (1.2, 3.1) {$x$} ;
      \draw [black, ->,dashed, thick] (1.2, 1.5) -- (0.4, 2.7) node [black] at (0.8, 2.7) {$\varphi$} ;
      \draw [black,thick,<-, domain=120:90] plot ({1.2*(1 + 0*sin(\x))*cos(\x) + 1.2}, {1.3 +  1.2*(1 + 0*sin(\x))*sin(\x)});
      \end{tikzpicture}}
       \caption{The focusing setup with focusing length $l$. We choose axis $x$ and $y$ such that $x$ is aligned along the axis of injection. We locate the
       source at $(x, y) = (0, 0)$. The in plane magnetic field angle $\phi$ is measured from the $x$ axis.}
          \end{center}
  \end{figure}
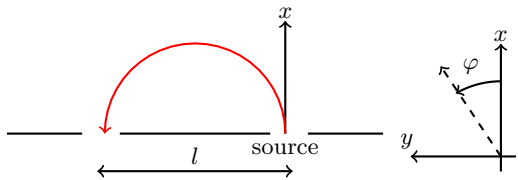

\begin{figure}[t!]
     {\includegraphics[width=0.42\textwidth]{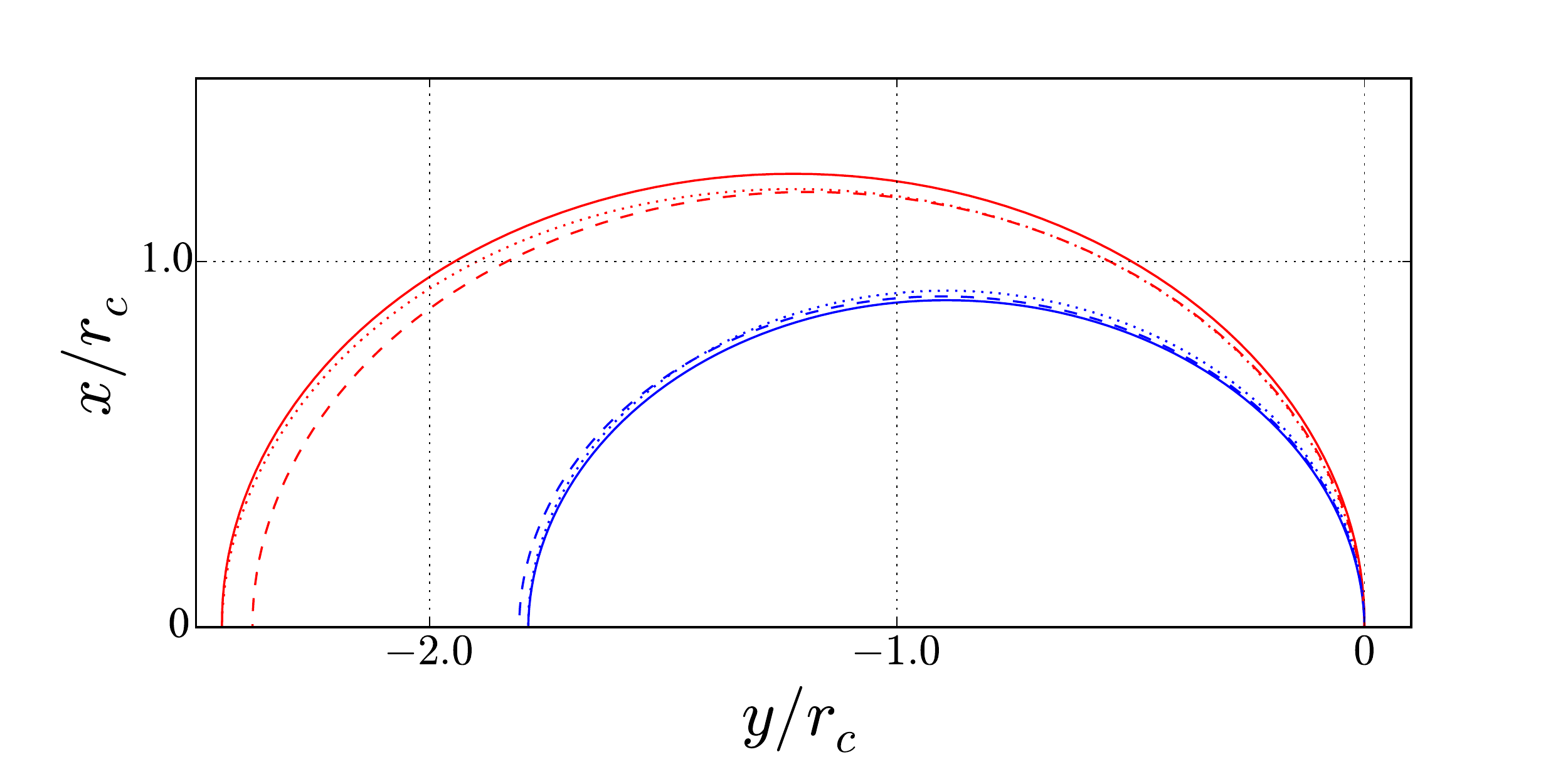}}
     {\includegraphics[width=0.42\textwidth]{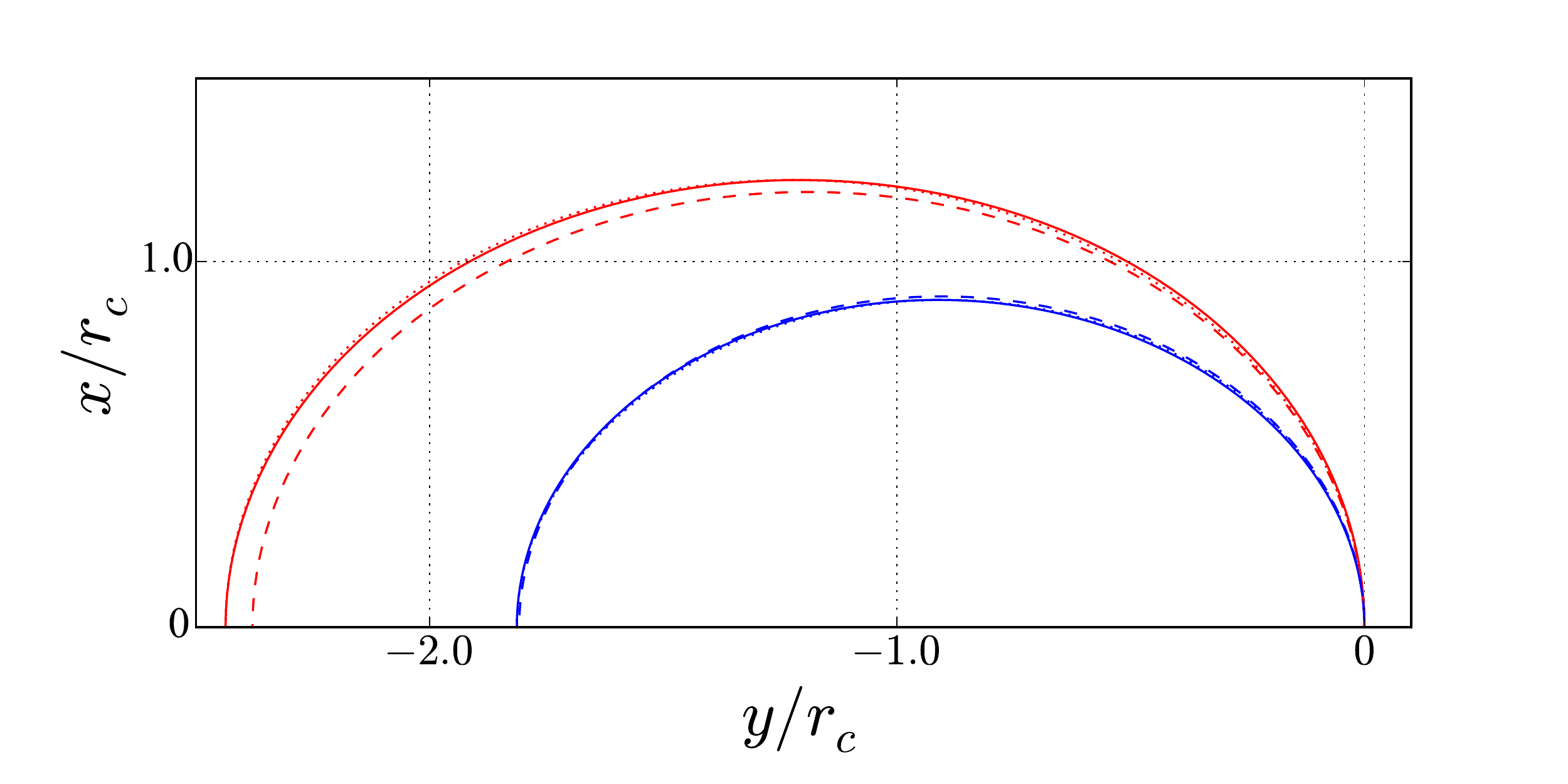}}     	
     \caption{Trajectories of spin-orbit coupled holes, with $s = 1$ in red, 
	and $s=-1$ in blue. We use $\tilde\gamma_3 = 0.25$ and $\tilde g_1 = 1$ with
	an in plane magnetic field, $B_{||} = 4$T. The Fermi energy is $\varepsilon_F = 1.9$meV. 
	The upper panel has a in plane magnetic field orientation $\phi = 0$, 
	while the lower panel has a field orientation $\phi = \pi/2$. We present the trajectories normalised with the
	cyclotron radii, with $r_c \approx l/2$. The dashed lines in the upper and lower panel have $B_{||} = 0$T. Note
	that the change in the trajectory is very small. We note that the injection with velocity directed fully along $x$ does 
	not correspond to $\theta_i = 0$.}
\label{trajectories}
\end{figure}

The equations of motion of this classical hamiltonian are
 \begin{eqnarray}
\label{heq}
&&v_+=\frac{\partial {\cal H}_{cl}}{\partial \pi_-}
=\frac{\pi_+}{m}-\frac{s}{|{\cal B}|}\frac{\partial {\cal B}^2}{\partial \pi_-}
\nonumber\\
&&{\dot {\pi}}_+=i\omega_c m v_+ \ ,
\end{eqnarray}
The solution to these classical equations of motion has been found with 
expansion in powers of $|{\cal B}|/\varepsilon_F$\cite{Bladwell2015}.  
The particle trajectories are given by
\begin{eqnarray}
\label{traj1}
&&\theta_0=\omega_ct\nonumber\\ 
&&\theta=\theta_0-s\frac{3|{\tilde \gamma_3}|}{2}\int_{\theta_i}^{\theta_0}
\frac{a(\theta^{\prime})}{b(\theta^{\prime})}d\theta^{\prime}\nonumber\\
&&x+iy=\frac{k_F}{m\omega_c}\left\{i(e^{i\theta_i}-e^{i\theta})\right.\nonumber\\
&&\left.\ \ \ \ \ \ \ \ \ +s\frac{|{\tilde \gamma_3}|}{2}\int_{\theta_i}^{\theta}e^{i\theta^{\prime}}
\left[b(\theta^{\prime})+3i\frac{c(\theta^{\prime})}{b(\theta^{\prime})}\right]
d\theta^{\prime}\right\}
\end{eqnarray}
where $k_F = \sqrt{2 m \epsilon_F}$ is the Fermi momentum. We have
introduced the initial angle $\theta_i$. 
The condition for the adiabatic evolution of the spin implies that $b(\theta)$ does not vanish,
$|{\tilde \gamma_3}| b(\theta)\gg \omega_c/\epsilon_F$.
The functions $c(\theta)$ and $a(\theta)$ are given by
\begin{eqnarray}
\label{tv}
&&{\dot \theta}=\omega_c\left[1-s\frac{3|{\tilde \gamma_3}|}{2}
\frac{a(\theta)}{b(\theta)}\right]\nonumber\\
&&a(\theta)=1+
(5/3)({\tilde g_1 \mu_B |{B_{||}}|}/{\tilde \gamma_3\varepsilon_F})\cos(\theta-\varphi) \nonumber \\
&&+(2/3)({\tilde g_1 \mu_B |B_{||}|}/{\tilde \gamma_3 \varepsilon_F})^2\nonumber\\
&&c(\theta)=\frac{1}{3} ({\tilde g_1 \mu_B |{B_{||}}|}/{\tilde \gamma_3 \varepsilon_F})\sin(\theta-\varphi) \ .
\end{eqnarray}
These solve the problem of the classical motion.
We have presented illustrative 
trajectories in Fig. \ref{trajectories}. The classical trajectories are essentially 
unchanged, even up to several Tesla. 

A peculiar feature to note is that $\theta_i=0$ does not 
correspond to the classical trajectory, since $\theta_i = 0$ has non-zero $v_y$. 
We define the physical injection angle, $\beta$ such that the classical trajectories shown in
Fig. \ref{trajectories} correspond to injection with $\beta = 0$. To relate this 
to $\theta_i$, we differentiate Eq. \eqref{traj1} to obtain $v_y$ at the source 
\begin{eqnarray}
v_y \approx &\frac{k_{F, s}}{m} \left(\sin\theta_i + \frac{\tilde g_1 \mu_B B_{||}}{2} \sin(2\theta_i + \varphi) \right) \\ \nonumber
&k_{F, s} = k_F \left(1 + s \frac{\tilde \gamma_3}{2 \varepsilon_F} \right) 
\end{eqnarray}
Setting $v_y =0$ and solving, we obtain $\theta_i  \approx \tilde g_1 \mu_B B_{||} \sin\varphi/2 \varepsilon_F$. In general, 
$\beta$ is related to $\theta_i$ by
\begin{eqnarray}
 \theta_i = \beta + \frac{\tilde g_1 \mu_B B_{||} \sin\varphi}{2 \varepsilon_F}
 \label{beta1}
 \end{eqnarray}
We stress again that the method used here, and following in 
Sec. \ref{interference} can equally be applied to electron systems with a Rashba SOI and 
an applied in-plane magnetic field. 

\section{Interference}
\label{interference}

The problem of interference in systems with 
large SOIs has been treated in detail for cylindrically symmetric systems\cite{Bladwell2017}. 
Like any interference problem, there are two trajectories (see Fig. \ref{int1}), connecting 
the source located at the origin, $(0, 0)$, to a detector located at $(0, l)$. These two paths are defined 
by injection angles $\pm \beta$, with
 \begin{eqnarray}
&\cos\beta = \frac{l}{2 r_{c, s}}   \\ \nonumber
&r_{c, s} = k_{F, s}/eB_z
\label{beta}
\end{eqnarray}
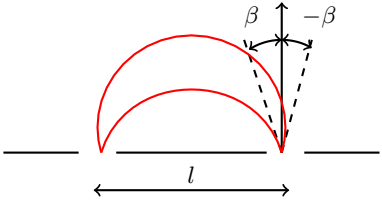
\begin{figure}[h!]
 \label{int1}
   \begin{center}
    {\begin{tikzpicture}
    \draw [black, thick] (-2.5, 1.5) -- (-1.5, 1.5);
    \draw [black, thick] (-1, 1.5) -- (1, 1.5);    
    \draw [black, thick] (1.5, 1.5) -- (2.5, 1.5);    
    \draw [black, thick, ->] (1.2, 1.5) -- (1.2, 3.5);    
      \draw [black, <->, thick] (-1.3, 1.0) -- (1.3, 1.0) node [black] at (0, 1.2) {$l$}; 
            \draw [black, dashed, thick] (1.2, 1.5) -- (1.6, 3.0) ;
                  \draw [black,thick,<->,domain=57:90] plot ({1.2 - 0.8*cos(\x)}, {2.2 + 0.8*sin(\x)}) node [black] at (1.7, 3.3) {$-\beta$};           
                   \draw [black,thick,<->,domain=90:120] plot ({1.2 - 0.8*cos(\x)}, {2.2 + 0.8*sin(\x)}) node [black] at (0.8, 3.3) {$\beta$};             
      \draw [black, dashed, thick] (1.2, 1.5) -- (0.7, 3.0) ;
                        \draw [red,thick,domain=0:180] plot ({1.2*(1 + 0.3*sin(\x))*cos(\x)}, {1.5 + 1.2*(1 + 0.3*sin(\x))*sin(\x)});
       \draw [red,thick,domain=0:180] plot ({1.2*(1 - 0.3*sin(\x))*cos(\x)}, {1.5 + 1.2*(1 - 0.3*sin(\x))*sin(\x)});
        \end{tikzpicture}}
       \caption{The two trajectories of injection angle $\beta$ and $-\beta$.}
          \end{center}
  \end{figure}
Interference
the arises from the difference between the phases of the two trajectories, with the 
semiclassical propagator defined as the sum over the two classically allowed paths, 
\begin{eqnarray}
K(\beta) \sim e^{i {\cal S}(\beta)} + e^{i {\cal S}(-\beta)}
\label{Kernel}
\end{eqnarray}
with the phase ${\cal S} \propto \int {\bm p}\cdot d{\bm l}$, where $d{\bm l}$ is integrated 
along the path of the trajectories. 
In a typical TMF setup, the source and detector are of some finite aperture, 
with the Huygens Kernel, Eq. \eqref{Kernel}
averaged over this aperture\cite{Bladwell2017}. 

Evaluation of the phase is treated analogously to the cylindrically symmetric case.  
The canonical is related to the kinematic momentum and the 
vector potential by ${\bm p} = m{\bm v} + e {\bm A}$, and the action is
\begin{eqnarray}
{\cal S} (\beta) = \int_{\delta S} \left(m{\bm v}  + e {\bm A} \right) \cdot d {\bm l} 
\label{phaseintegral}
\end{eqnarray}
where ${\bm v}$, ${\bm A}$, $d{ \bm l}$ and the path, $\delta S$, are dependent on $\tilde g_1 \mu_B B_{||}$. 
Using the previously determined equations of motion,
 Eqs. \eqref{traj1} and \eqref{tv}, the phase integral, Eq. \eqref{phaseintegral} can converted 
 into an integral over the running angle, 
  \begin{eqnarray}
{\cal S} (\beta) = eB_z \int_{\theta_i}^\theta \left(\left(\frac{dx}{d\theta'}\right)^2  + \left(\frac{dy}{d\theta'}\right)^2 - x \frac{dy}{d\theta'} \right) d {\theta'} 
\label{phaseintegral2}
\end{eqnarray}
The relationship between the physical injection angle $\beta$ and 
$\theta_i$ is presented in Eq. \eqref{beta1}. We must also determine 
$\theta$ in terms of $\beta$.

The trajectory 
from the source to the detector is, in terms of the running angle, from $\theta_i$ to $\theta$. 
This corresponds to the spatial positions $(0, 0)$ and $(0, l)$ respectively. From Eq. \eqref{traj1} we have
\begin{eqnarray}
x = 0 = &&\frac{k_s}{\omega_c m} \big\{(\sin\theta_i + \sin\theta) \\
&&+ s\frac{\tilde g_1 \mu_B B_{||}}{4} \left[ \sin(2\theta - \varphi) - \sin(2\theta_i - \varphi)\right] \big\} \nonumber.
\label{xe0}
\end{eqnarray}
We have restricted ourselves to a first order expansion in $\tilde g_1 \mu_B B_{||}$ when 
performing the integration of Eq. \eqref{traj1}. 
The trajectory deviates only minimally from the arc of a circle (see Fig. \ref{trajectories}), and we can reasonably 
employ the approximation
$\theta \approx \pi - \theta_i$ for the $\tilde g_1 \mu_B B_{||}$ dependent terms. With this approximation, 
solving Eq. \eqref{xe0} we obtain
\begin{eqnarray}
\theta \approx \pi - \theta_i  - {s \tilde g_1\mu_B B_{||}}\sin\theta_i \cos\varphi 
\label{angleoffset2}
\end{eqnarray}
Finally, this can be expressed in terms of the injection angle, $\beta$ using Eq. \eqref{beta1}, to obtain 
the integration limits for Eq. \eqref{phaseintegral2} in terms of $\beta$.

Using these integration limits, integration
of Eq. \eqref{phaseintegral2} yields
\begin{eqnarray}
&S(\beta) = \frac{k_s^2}{2eB} \Big\{ \pi - 2\beta + \sin2\beta\nonumber + \zeta \\ \label{phases}
& - s {\tilde g_1} \mu_B B_{||} \sin\beta (1 - \cos2\beta) \cos\varphi \Big\} \\ \nonumber
&\zeta = {\tilde g_1} \mu_B B_{||} \Big\{\sin\varphi - \cos2\beta\sin\varphi + \sin\varphi(\cos\beta + \frac{1}{3} \cos^3\beta) \\ \nonumber
&\cos\varphi(\cos\beta + \frac{1}{3} \cos^3\beta) \Big\}
\label{phase}
\end{eqnarray}
For $-\beta$ injection angles, we take $\beta \rightarrow -\beta$.
We have introduced here $\zeta$ which 
contains the phase terms that do not contribute any net phase difference,
that are symmetric for $\beta \rightarrow -\beta$.
According to Eq. \eqref{Kernel}, we then have
\begin{eqnarray} 
&K(\beta) \sim e^{i {\cal S}(\beta)} + e^{i {\cal S}(-\beta)} \\ \nonumber
&\sim \sin\Big[\frac{k_s^2}{2eB} \Big(2\beta -  \sin2\beta  \\ \nonumber
&+ s {\tilde g_1} \mu_B B_{||} \sin\beta (1 - \cos2\beta) \cos\varphi \Big) + \frac{\pi}{4} \Big]
\label{Kernel2}
\end{eqnarray}
The additional factor of $\pi/4$ arises due to the caustic for the $-\beta$ path\cite{Bladwell2017}.
 The third line of 
Eq. \eqref{Kernel}, which is linear in $\tilde g_1\mu_B B_{||}$, represents the ``emergent phase contribution'', 
and is the first major result of this work. 
This term is particularly remarkable, since the
classical trajectories have no first order dependence on $\tilde g_1 \mu_B B_{||}$.
For quantum 
(Shubnikov de Haas) oscillations, the integral is over the entire Fermi surface, and this term vanishes. Thus 
it is peculiar to the particular geometry of TMF, which defines the angle, $\varphi$ between the in plane magnetic
field and the injector. 

\begin{figure}[t!]
     {\includegraphics[width=0.42\textwidth]{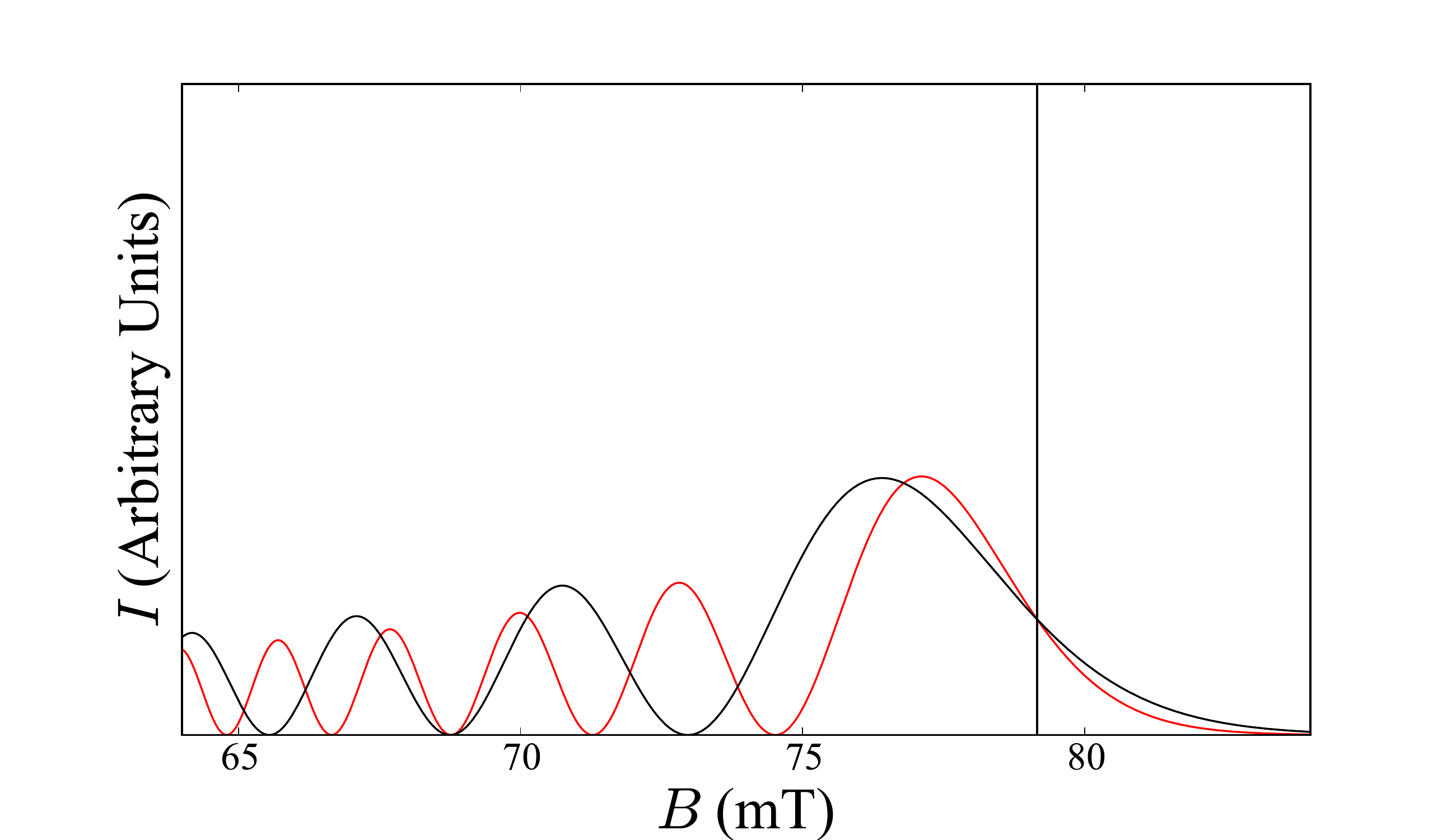}}
     \caption{Interference pattern calculated for point-like source and detector from Eq. \eqref{kso1}. We use 
	$l = 1500$nm, $k_F = 0.107\times10^{-1}$nm$^{-1}$ with a Rashba splitting $\tilde\gamma_3= 0.3\varepsilon_F$, 
	and $\tilde g_1 \mu_B B_{||} = 0.15\varepsilon_F$, corresponding to $B_{||} \sim 4-5$T for $\tilde g_1 = 1$ and 
	$\varepsilon_F \approx 2$meV. For clarity we present only a single spin state.
	The Interference spectrum is calculated from Eq. \eqref{kso1}, with Black vertical line indicates the location of the classical cutoff. 
	Red (black) plots have  in-plane field with $\varphi = \pi$ ($\varphi = 0$)}
\label{Fig1}
\end{figure}

It is instructive to 
examine the variation in the interference fringe separation due to the 
application of the in magnetic plane field, expanding for small $\beta$. 
For small $\beta$, according to Eq. \eqref{beta},
\begin{eqnarray}
\beta \approx \sqrt{\frac{2r_{c, s} - l}{r_{c, s}} } = \sqrt{\frac{y}{r_{c, s}}}
\end{eqnarray}
Here $y = 2r_{c, s} - l$ is the detuning from the classically forbidden region. 
For small $\beta$, Eq. \eqref{phase} becomes
\begin{eqnarray}
{\cal S} \approx \frac{2}{3} \nu_s \left(\frac{y}{r_c}\right)^\frac{3}{2} \left(1  +  s\frac{3}{2} \tilde g_1 \mu_B B_{||} \cos\varphi  \right)
\label{phase2}
\end{eqnarray}
And from Eq. \eqref{Kernel2},
we find a characteristic spacing of the interference fringes to be
\begin{eqnarray}
\frac{\delta B}{B} \approx \frac{2.2}{2 \nu^\frac{2}{3}} \left(1 - s \tilde g_1 \mu_B B_{||} \cos\varphi  \right)
\label{fringeseparation}
\end{eqnarray}
where $\delta B$ is the fringe spacing. 
This provides a method of determining the 
strength of the secondary spin orbit interaction. As can be seen in Fig. \ref{Fig1}, even for the 
first interference fringe, there is a measurable shift.
While there is no direct enhancement, 
the strength of the $\tilde g_1 \mu_B B_{||} \sim 0.1\varepsilon_F$ at fields of a few Tesla in hole 
systems.
Recent TMF experiments
have resolved a single interference fringe for the low field peak\cite{Rokhinson2004, Lo2017}, 
which would sufficient for the determination of $\tilde g_1 \mu_B B_{||}$. 

The remaining elements of the Huygen's kernel are unchanged cyclindrically 
symmetric case. As was detailed in Ref. [\cite{Bladwell2017}], the asymptotic form of the Huygen's kernelcan be related to the Airy function. 
Employing the same reasoning, from
Eqs. \eqref{Kernel2} and \eqref{phase2}, we obtain, 
 \begin{eqnarray}
\label{kso1}
 K_s =&& e^{i\frac{\pi (\nu_s-1-n)}{2}}\frac{\nu_s^{2/3}}{2\sqrt{2}r_{cs}} \times \\ \nonumber 
&& \left[(\sigma_z-is \sigma_x)Ai(\overline{y}_s) +\frac{n}{\nu_s^{1/3}}Ai^{\prime}(\overline{y}_s)\right] \ .
 \end{eqnarray} 
 Here $\overline{y} = y \nu_s^{2/3} \left(1 + s \tilde g_1 \mu_B B_{||} \cos\varphi\right)/r_c$. 
We present plots of the resulting interference spectrum in Fig. \ref{Fig1} with point-like sources and 
detectors, for both the classical form of the Huygen's kernel, and Eq. \eqref{kso1}.
We stress that this semi-classical approach employed is only valid if $\nu_s \gg 1$. For typical
experimental systems, $\nu_s > 30$.

\section{Discussion}
\label{discuss}

\begin{figure}[t!]
     {\includegraphics[width=0.42\textwidth]{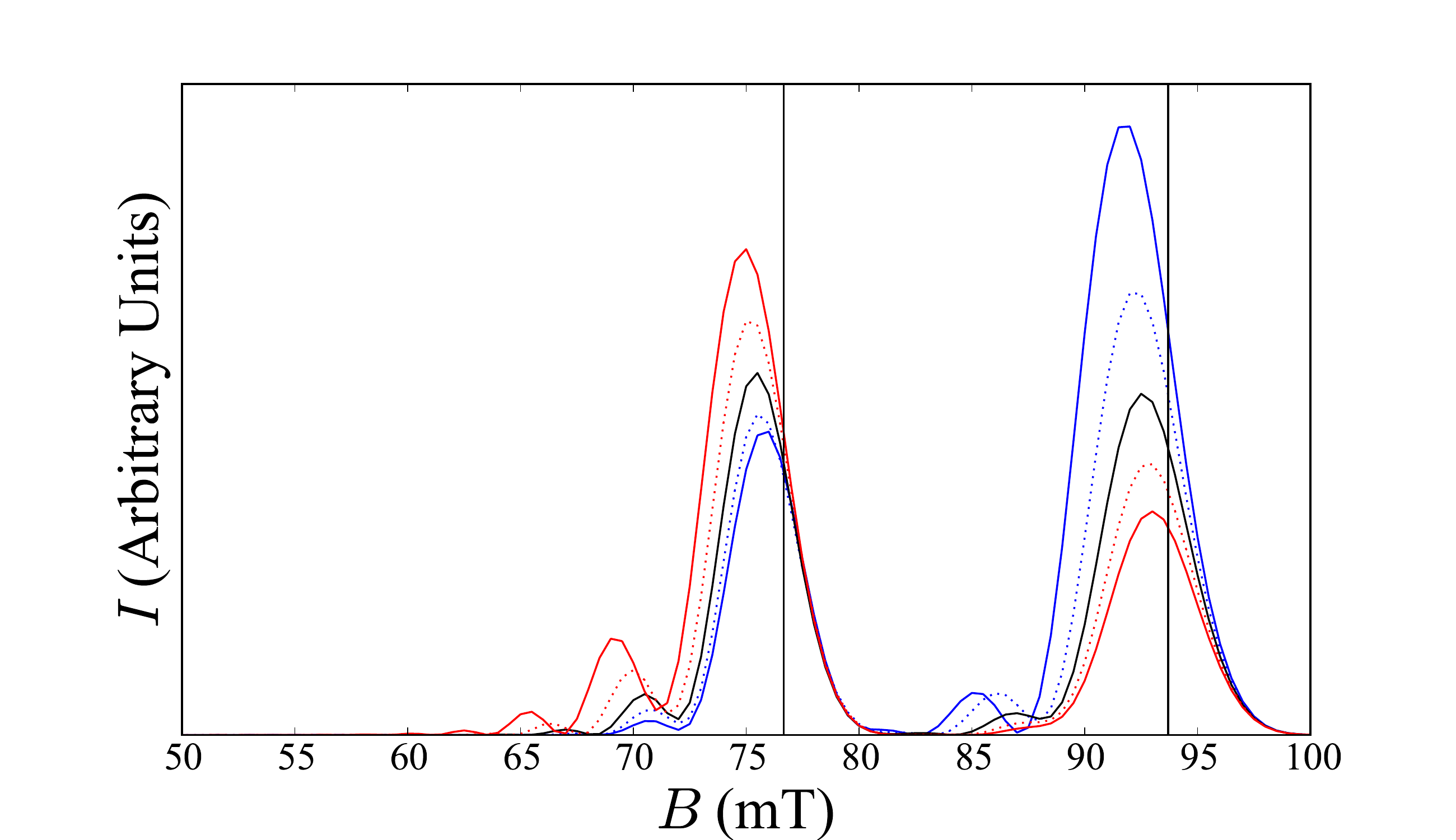}}
     \caption{Interference patterns versus $\tilde g_1 \mu_B B_{||}$, with range $-0.1< \tilde g_1 \mu_B B_{||} < 0.1$, with
     $\varphi = 0$. Here
	$l = 1500$nm, $w= 150$nm. Positive $\tilde g_1 \mu_B B_x$ is presented in red, while negative $\tilde g_1 \mu_B B_x$
	are in blue. The $B_x= 0$ TMF spectrum is solid black curve. Vertical black lines indicate the location of the classical maximum, 
	$B_{focusing} = 2 \hbar k_s/el$.
}
\label{FigParallel}
\end{figure}

In real systems, the source and detector have finite size and can influence the 
observed interference pattern. Typically experimental
devices use quantum point contacts, which consist of a narrow channel connecting 
a reservoir to the 2DHG. These have some characteristic width, $w$, and 
can be modelled by standing waves in the 
$y$ direction,
\begin{eqnarray}
\label{psisd}
\psi_{s} &\propto& \chi_s \sin\left(\frac{\pi y}{w} \right)  \ \ \ \ \ \ \  \ \ \ \ \  0 < y < w\nonumber\\
\psi_{d} &\propto& \chi_d \sin\left(\frac{\pi (y-L)}{w} \right)  \ \ \  L < y < L+w \ .
\end{eqnarray}
where $w$ is the width of the channel, and $\chi_s$ and $\chi_d$ are the 
eigenspinors at the source and detector respectively\cite{Bladwell2017}. 
The 
exit width, $w$, is imposed by the lithographic geometry of the
QPC\cite{Molemkamp1990}, however can vary depending on the 
conductance. 
We consider a hole gas with a density, $n=1.85 \ 10^{11}cm^{-2}$, and corresponding 
Fermi momentum, $k_F = 0.107nm^{-1}$. With a Rashba splitting 
of $\tilde\gamma_3 = 0.2 \varepsilon_F$, 
the smaller spin-split Fermi momentum will be $k_-=0.096nm^{-1}$ ($\lambda \approx 66nm$)
and the larger Fermi momentum $k_+=0.118nm^{-1}$ ($\lambda \approx 53nm$).
The distance between the source and the detector is $L=1500nm$.
The corresponding magnetic field at the  classical edge of the bright region for $k_-$
is $B_{-}=77mT$, while for $k_+$, $B_{+}=94mT$.
We will start by considering a QPC of width $w = 150$nm, 
which we note corresponds to the lithographic width of 
the device of Ref \cite{Rokhinson2004}. 
The resulting focusing spectrum is presented in Fig. \ref{FigParallel}. 
While interference fringes are still visible, they are suppressed, 
and the additional phase contribution manifests as a suppression or 
enhancement of the spin-split focusing peaks. We note that this experimental 
signature is similar to that typically attributed to polarisation in the
QPCs\cite{Reynoso2007}.

In summary, we have employed Huygen's principle to determine 
quantum interference for systems with asymmetrical Fermi surfaces.
While in this work we focus on a specific case of an in-plane magnetic 
field in combination with a Rashba spin-orbit interaction, the method 
employed is general. We have predicted an emergent phase contribution, 
linear in the applied in plane magnetic field, despite there being no first 
order changes to the classical trajectories. 
This emergent phase term significantly alters the interference spectrum of
TMF. We propose that this could be used to measure the in plane $g$ factor.

\section{Acknowledgements}

We thank Alex Hamilton, Dmitry Misarev, Harley Scammell and 
Matthew Rendell for their helpful discussions. SSRB 
acknowledges the support of the Australian Postgraduate 
Award. This research was partially supported by the 
Australian Research Council Centre of Excellence in Future Low-Energy Electronics Technologies
 (project number CE170100039) and funded by the Australian Government.

\end{document}